\documentclass[9pt,twocolumn,twoside]{osajnl}
\journal{ol}
\setboolean{shortarticle}{true}

\usepackage{comment}
\usepackage{txfonts}
\usepackage{amsmath,amssymb}
\usepackage{graphicx}
\usepackage{upgreek}
\usepackage[detect-weight]{siunitx}
\DeclareSIUnit\bar{bar}
\usepackage{soul,color}

\title{Supercontinuum generation in methane-filled hollow-core antiresonant fiber}

\author[1]{Balazs Plosz}
\author[1]{Athanasios Lekosiotis}
\author[1]{Mohammad Sabbah}
\author[1]{Federico Belli}
\author[1]{Christian Brahms}
\author[1,*]{John C. Travers}

\affil[1]{School of Engineering and Physical Sciences, Heriot-Watt University, Edinburgh, EH14 4AS, UK}

\affil[*]{Corresponding author: j.travers@hw.ac.uk}

\begin{abstract}
We report the generation of a multi-octave supercontinuum spanning from 350 nm to 1700 nm with exceptional spectral flatness and high conversion efficiency to both the visible and near infrared region, by pumping a methane-filled hollow-core antiresonant fiber with 1030 nm laser pulses. The dynamics exhibited signs of both modulational instability and stimulated Raman scattering. Fiber lengths ranging from 15 to 200~cm were investigated along with gas pressures up to 50 bar and pump pulse durations from 220~fs up to 10~ps. The best supercontinuum, in terms of spectral width and flatness, was achieved with 220~fs pulses, 25~bar filling pressure, and 60~cm propagation length. Comparison with argon-filled fiber with matched nonlinearity and dispersion showed that the Raman contribution enhances the supercontinuum generation process compared to a pure modulational instability-based process. The average power was scaled up by increasing the pulse repetition rate to 50~kHz, but further scaling was hindered by linear and nonlinear absorption leading to fiber damage.
\end{abstract}

\setboolean{displaycopyright}{true}

\begin{document}

\maketitle

\noindent Gas-filled hollow-core fibers have proven to be a very effective system for nonlinear optics, such as frequency conversion, supercontinuum generation, and optical pulse compression~\cite{tani_phz-wide_2013,belli_vacuum-ultraviolet_2015,ermolov_supercontinuum_2015, Debord_2015, mak_tunable_2013,travers_high-energy_2019,nisoli_generation_1996}. Hollow-core antiresonant fibers in particular offer the benefit of low loss and high transmission bandwidth with strong mode confinement \cite{travers_ultrafast_2011}, enabling efficient broadband supercontinuum generation~\cite{tani_phz-wide_2013,belli_vacuum-ultraviolet_2015,ermolov_supercontinuum_2015,gao_raman_nodate,Sabbah:23}. This is typically achieved by pumping in the anomalous dispersion region to access modulational instability~(MI)~\cite{tani_phz-wide_2013} or soliton self-compression dynamics~\cite{belli_vacuum-ultraviolet_2015,ermolov_supercontinuum_2015}.  When pumping noble gases in the normal dispersion region, the spectral broadening is driven mostly by self-phase modulation. This offers only moderate spectral broadening due to dispersion. Attempts to overcome this by increasing the peak power are limited by self-focusing and plasma effects.

The use of molecular gases provides additional nonlinearity through the excitation of the rotational and vibrational degrees of freedom of molecules through stimulated Raman scattering (SRS)---an effect well established in solid-core nonlinear fiber optics, where Raman scattering is always present and greatly enhances supercontinuum formation~\cite{dudley_supercontinuum_2006}. While Raman scattering in gas-filled hollow-core fibers has also been widely investigated \cite{HannaD.1986SRso,benabid_stimulated_2002,WangZefeng2014E1e,Gladyshev_2019,beetar_thermal_2021,ArcosPau2024SRSa}, it has not been greatly exploited for supercontinuum generation. The first supercontinuum generation in molecular gas-filled antiresonant fiber made use of Raman-enhanced soliton self-compression in the anomalous dispersion regime, producing a continuum extending to the vacuum ultraviolet in hydrogen and deuterium~\cite{belli_vacuum-ultraviolet_2015}. The use of atmospheric air filled fiber has also been considered~\cite{Debord_2015, mousavi_nonlinear_2018} in the picosecond pump regime.

Recently, a flat and broadband supercontinuum was generated by pumping nitrogen-filled antiresonant fiber in the normal dispersion regime, at 532~nm~\cite{gao_raman_nodate}. The mechanism in that work was identified to involve two stages: (i) the initial formation of a broad vibrational Raman frequency comb; (ii) the smoothing of the comb lines into a flat supercontinuum through Kerr-driven self-phase modulation and rotational Raman broadening. We name this mechanism comb-to-continuum. Using that approach, a supercontinuum spanning from \SI{440}{\nm} to~\SI{1200}{\nm} in a nitrogen-filled fiber was achieved~\cite{gao_raman_nodate}. We note that this approach builds on earlier frequency-comb work in gas-filled fibers~\cite{tani_generation_2015}, work in liquid-filled capillaries~\cite{Fanjoux:17}, and work in solid-core fibers~\cite{Mussot:03}. The comb-to-continuum technique is attractive because the extent of the supercontinuum is established by the initial Raman frequency comb, avoiding the need to broaden directly from the pump wavelength. This suggests that extremely broadband supercontinua could be generated. At first sight, nitrogen appears to be the ideal gas medium because its rotational Raman lines are very closely spaced, which aids in smoothing the vibrational frequency comb into a supercontinuum~\cite{gao_raman_nodate}. However, it was recently found that the closely spaced rotational lines lead to universal Raman gain suppression, which transfers energy to higher-order modes and reduces the efficiency of supercontinuum formation~\cite{Sabbah:23}.

In the current work, we investigated the use of methane (CH$_4$) for supercontinuum generation in antiresonant fibers in the anomalous dispersion regime, where the Raman dynamics interact with modulational instability arising from the Kerr nonlinearity. Methane is a symmetric molecule and, therefore, has no rotational Raman response. As a result, gain suppression due to closely spaced rotational lines is inherently avoided. In contrast to previous work on pumping methane-filled antiresonant fibers~\cite{Gladyshev_2019,Lanari:23}, we used a fiber with very thin core walls (\qty{147}{\nm} thickness), which place the first-order high-loss resonance band around \qty{313}{\nm}. This allows the supercontinuum (pumped at \qty{1030}{\nm}) to fully form in the fundamental guidance band. In addition, we compared the supercontinuum formation dynamics between argon and methane-filled fibers to identify the role of the Raman response on the flatness of the supercontinuum, and investigated the effect of power scaling on the formation of the supercontinuum.

\begin{figure}[tb]
\centering
\includegraphics[width=1\linewidth]{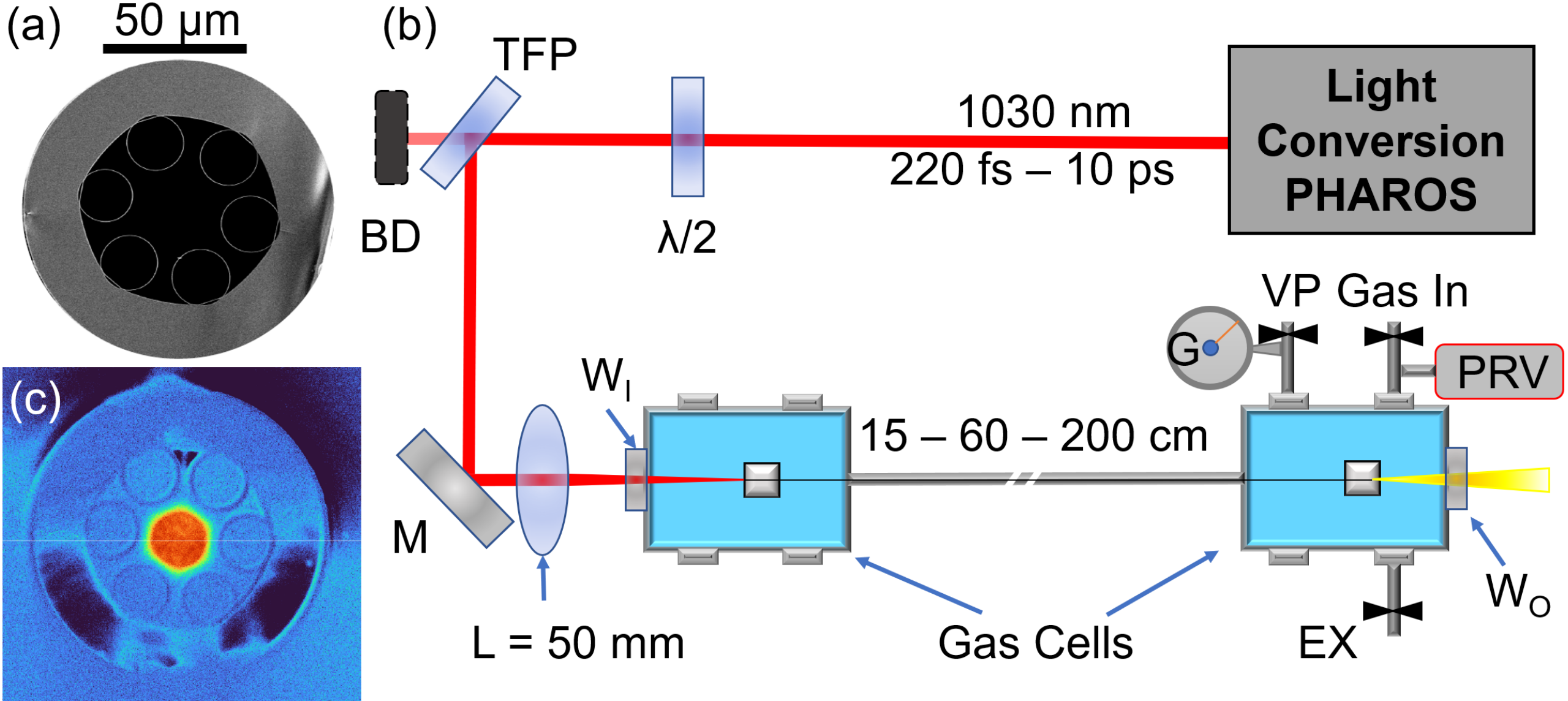}
\caption{(a) SEM image of the fiber (b) Sketch of the experimental setup. TFP: thin-film polarizer. $\uplambda /2$: half-wave plate. BD: beam-dump. M: mirror. {$\mathrm{W}_\textnormal{I}$}: input window. $\mathrm{W}_\textnormal{O}$: output window. G: pressure gauge. VP: vacuum pump line. Gas In: gas input. EX: exhaust line. PRV: pressure relief valve. (c) Near field output of the fiber at low pulse energy.}
\label{fig:setup}
\end{figure}

The hollow-core fiber we used was a single-ring nodeless antiresonant fiber with a core diameter of \qty{32}{\um} and wall-thickness of \qty{147}{\nm}. Fig.~\ref{fig:setup}(a) shows a scanning electron micrograph (SEM) of the antiresonant fiber cross-section, and Fig.~\ref{fig:setup}(c) shows the near-field output of the fiber recorded through a microscope. The experimental setup is shown in Fig.~\ref{fig:setup}(b). A Light Conversion PHAROS laser provided the pulses centered at 1030~nm with 220~fs FWHM transform-limited pulse duration. A thin-film polarizer (TFP) and half-wave plate ($\uplambda /2$) were used for power control. The beam was coupled into the fiber with a single 5~cm focal length anti-reflection coated plano-convex lens though a 6.3~mm thick anti-reflection coated fused silica window. The two gas cells, where the input and output of the fiber were located, were connected by a steel tube and the whole setup was designed to support gas pressures up to \qty{50}{\bar}. The output cell had a \qty{6.3}{\mm} thick uncoated fused silica window for broadband transmission. The output spectrum was collected using an integrating sphere and two fiber-coupled spectrometers (Avantes ULS2048XL, Avantes NIR256-2.5), which combined cover the range from 200 to \qty{2500}{\nm}. The whole system, starting from the integrating sphere, was calibrated on an absolute scale with NIST-traceable lamps.

In our experiments we systematically investigated different fiber lengths between \qty{15}{\cm} and \qty{200}{\cm}, different pump pulse durations between \qty{200}{\fs} and \qty{10}{\ps} (using the laser's built-in pulse compressor to chirp the pulses), and different gas pressures between \qty{5}{\bar} and \qty{50}{\bar}. Here we only describe the best results. Unless otherwise specified, all results shown here were obtained at \qty{1}{\kHz} repetition rate. Scaling of the repetition rate is described at the end of this letter.

Methane has a dominant vibrational mode at a frequency of $\Omega\approx 2\pi\times \qty{87}{\THz}$~\cite{bermejoAbsoluteRamanIntensities1977}, corresponding to an oscillation period of \qty{11.4}{\fs}, with a dephasing time $T_2$ above \qty{10}{\ps} for all the pressures we consider. Therefore, all our experiments are in the transient regime, in which the refractive index change approximately follows the pulse during the initial stages of the dynamics~\cite{belli_vacuum-ultraviolet_2015}.

\begin{figure}[tb]
\centering
\includegraphics[width=1\linewidth]{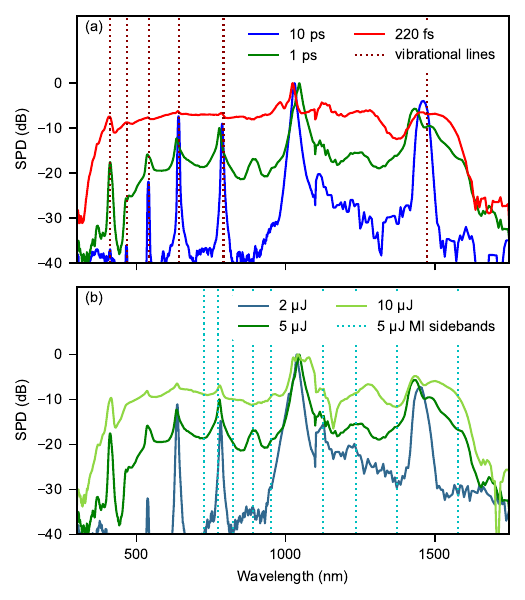}
\caption{Experimental spectra at the output end of 60 cm antiresonant fiber filled with 25 bar methane. (a) Results with \qty{5}{\uJ}, \qty{1030}{\nm} pulses with various input pulse durations; 10 ps shown in blue ($N=194$, $P_\mathrm{peak} =  \qty{0.44}{\MW}$), 1 ps in green ($N=61$, $P_\mathrm{peak} =  \qty{4.4}{\MW}$) and 220 fs in red ($N=29$, $P_\mathrm{peak} =  \qty{20}{\MW}$). The Raman vibrational lines are shown with dark red dashed lines. The $v_1$ mode with \qty{87}{\THz} Raman frequency shift puts the first Stokes emission from the 1030~nm pump at 1473~nm and the second at 2580~nm. The latter is, however, outside the guiding window of the fiber itself. The anti-Stokes lines are at 793~nm, 644~nm, 542~nm, 468~nm, 412~nm, 367~nm respectively up to the 6th order. (b) The 1~ps case with various pulse energies. This shows similar dynamics to changing the pulse duration. The first few MI sideband locations are shown for the 1~ps case with dotted lines.}
\label{fig:10ps_1ps_220fs_subplots}
\end{figure}

Fig.~\ref{fig:10ps_1ps_220fs_subplots} (a) shows the spectra achieved with three different pulse durations (\qty{10}{\ps}, \qty{1}{\ps} and \qty{220}{\fs}) for a \qty{60}{\cm} long fiber filled with \qty{25}{\bar} methane. The pump pulse wavelength lies in the anomalous dispersion regime. For the parameters shown, the soliton order~$N$ was sufficiently large that the electronic response would drive modulational instability dynamics, rather than soliton fission~\cite{dudley_supercontinuum_2006} ($N=194$, $N=61$, $N=29$ for the three pulse durations).

In Fig.~\ref{fig:10ps_1ps_220fs_subplots} (b) we can see the \qty{1}{\ps} case with various energies (the \qty{1}{\ps} \qty{5}{\uJ}, green color is the same in both (a) and (b)). We obtain a similar evolution to the case of changing pulse duration and can observe the ``comb-to-continuum'' dynamics previously observed in Ref.~\cite{gao_raman_nodate}.  For the \qty{1}{\ps} pump pulse duration, it is clear that both MI sidebands and a Raman frequency comb are formed over the \qty{60}{\cm} length scale. Only the Raman comb can be observed for the \qty{10}{\ps} pump pulse duration. This is because the lower peak power is insufficient for broadening of the comb lines to occur within the \qty{60}{\cm} fiber length. In the case of the \qty{220}{\fs} pump pulses (shown in Fig.~\ref{fig:10ps_1ps_220fs_subplots}(a)) we see a fully developed flat supercontinuum with very small peaks remaining at the Raman vibrational frequencies.

\begin{figure}[tb]
\centering
\includegraphics[width=0.93\linewidth]{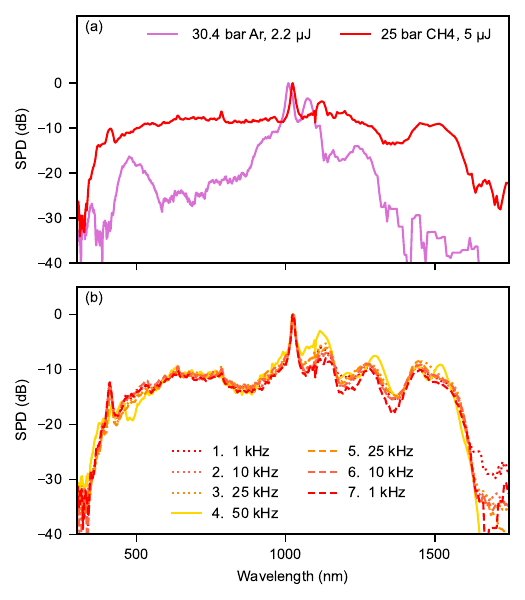}
\caption {(a) Supercontinuum spectra generated with either 25 bar of methane (red) or 30.4 bar of argon (pink) in the same, 60 cm, antiresonant fiber and pumped with \qty{220}{\fs} pulses with \qty{5}{\uJ} in the case of methane and \qty{2.2}{\uJ} in the case of argon. These parameters match the zero dispersion point at \qty{816}{\nm} and pump soliton order of 29. The spectra are normalized to their peaks. (b) Repetition rate scaling for \qty{25}{\bar} methane in the \qty{60}{\cm} fiber, pumped with \qty{220}{\fs}, \qty{5}{\uJ} pulses at \qty{1030}{\nm}. The numbered lines indicate the order of repetition rate variation. The maximum repetition rate with no damage was \qty{50}{\kHz}.}
\label{fig:Ar_vs_CH4_reprate}
\end{figure}

Fig.~\ref{fig:Ar_vs_CH4_reprate} (a) compares the methane results with argon. We adjusted the gas pressure to match the zero dispersion wavelength and the pump energy to match the soliton order $N$. It is clear that apart from the initial side-band formation and dispersive wave emission around \qty{500}{\nm}, the final spectral broadening is much less developed in argon than methane. This is in distinct contrast to the comparison between nitrogen and argon in Ref.~\cite{Sabbah:23}, where the universal gain suppression due to the closely spaced rotational lines in nitrogen reduced the total spectral power density, making the supercontinuum less established than Raman-free argon. Our work here shows that when carefully avoiding gain suppression---by avoiding closely spaced rotational lines through the use of methane---the additional nonlinearity from the vibrational Raman response actually enhances the supercontinuum. 

In the methane case, pumping with \qty{5}{\uJ}, we achieved slightly over $20\%$ conversion efficiency to both the short-wavelength and long-wavelength regions (below $\qty{1000}{\nm}$ and above $\qty{1060}{\nm}$) with good spectral flatness within 10~dB of the peak of the spectrum. The conversion saturated above this pump energy level.

Power-scaling of supercontinuum sources is of great importance to many applications, such as industrial metrology, where the system throughput is directly related to spectral power per unit area. All the results above were taken at \qty{1}{\kHz} pulse repetition rate. We performed a series of experiments to test the power scaling of methane-based supercontinuum generation.

First, we tested repetition rate scaling. The results are shown in Fig.~\ref{fig:Ar_vs_CH4_reprate}(b). In these experiments, the exposure of light into the fiber lasted just a few seconds in total, with the repetition rate being first stepped up and then back down---from \qty{1}{\kHz} to \qty{50}{\kHz} then back to \qty{1}{\kHz}---in quick succession. These data show that there is only moderate repetition rate dependence of the supercontinuum dynamics up to \qty{50}{\kHz}.

Second, we tested the lifetime of the system at \qty{10}{\kHz} repetition rate, corresponding to just \qty{50}{\mW}, but for a longer period of time. Even at this low repetition rate, transmission through the fiber was lost after just a few minutes due to permanent fiber damage. The damage to the fiber did not occur at the input facet, as would be expected if it were caused by pump beam pointing instability. Instead, from detailed examination of side-on microscope images of the fiber, it was clear that the degradation occured several hundred micrometers inside the fiber, from the input end. The appearance of damage inside the fiber was surprising since no similar effect was observed with other gases at such a low power, including argon, nitrogen, or hydrogen, to our knowledge, and antiresonant fibers have been shown to handle extreme intensities~\cite{Lekosiotis:23}. At the position of the damage, no significant spectral broadening of the pump light has occurred, therefore, it is clear that the damage mechanism must be due to the interaction of the pump light with the gas.

We hypothesize that the damage is due to temperature increases caused by both linear and nonlinear absorption in the filling gas. Methane is very sensitive to temperature increases and starts to decompose in 3 ways~\cite{cantelo2002thermal}: $2\mathrm{CH}_4 \rightarrow \mathrm{C}_2\mathrm{H}_4 + 2\mathrm{H}_2$, $2\mathrm{CH}_4 \rightarrow \mathrm{C}_2\mathrm{H}_2 + 3\mathrm{H}_2$, and $\mathrm{CH}_4 \rightarrow \mathrm{C} + 2\mathrm{H}_2$. The latter reaction dominates the process, and a negligible amount of ethylene or acetylene can be expected in the system. This decomposition starts to happen at a few hundred degrees, and above several hundred degrees virtually no methane should be present~\cite{cantelo2002thermal}. This idea was supported by the observation of black dust inside the gas cell when using higher repetition rates, which we believe to be the carbon produced by the decomposition of methane. A similar effect has previously been observed before using ethylene, adding further support to our hypothesis \cite{haddad_molecular_2018}. Furthermore, this idea can also explain why the damage occurs inside the fiber rather than at the fiber input facet, where convection of the gas will reduce the temperature increase.

\begin{figure}[tb]
    \centering
    \includegraphics[width=1\linewidth]{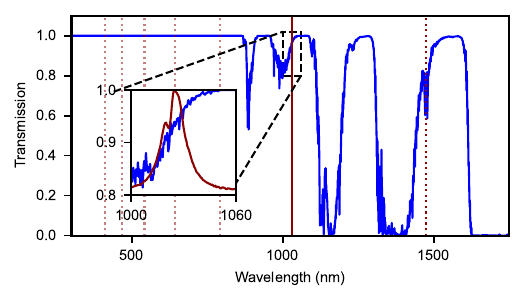}
    \caption{Calculated linear transmission of 25 bar methane at \qty{20}{\degreeCelsius} through a 1~m length (blue). The experimental pump pulse input is shown with a dark red line, and the Raman vibration lines are shown with dashed lines. The inset shows the transmission around the pump spectrum. This data is acquired from the high-resolution transmission molecular absorption database~\cite{GORDON2022107949}.}
    \label{fig:transmission}
 \end{figure}

To estimate the feasibility and time required for the methane to reach its decomposition temperature in our experiment, we make several significant simplifying assumptions: we neglect gas diffusion and conduction, and assume constant pressure. Based on the constant pressure heat capacity of methane at \qty{20}{\degreeCelsius} of $c_p=\qty{35}{\J\per\mol\per\K}$~\cite{GurvichAV1967TPOI}, and the molar volume of methane for our conditions (pressure $p=\qty{25}{\bar}$, temperature $T=\qty{20}{\degreeCelsius}$) $V_m=\qty{9.3e-4}{\m^3\per\mol}$, we can estimate the temperature rise as a function of energy absorption per unit length $\delta E/\delta z = \alpha E_\mathrm{in}$ at position $z=0$, with linear absorption coefficient $\alpha$ and pump pulse energy $E$, to be
\begin{equation}
\label{eqn:dt}
\Delta T = \alpha E V_m / c_p A
\end{equation}
where $A$ is the cross-sectional area of the fiber core (we assume a circle with radius equal to the inscribed core-size of the antiresonant fiber, $r=\qty{16}{\um}$). Hydrocarbons like methane often have linear near-infrared absorption. As shown in Fig.~\ref{fig:transmission}, our pump spectrum overlaps with a strong absorption region in methane. Furthermore, the final supercontinuum spectrum reaches into regions of very strong absorption in the near-infrared. Some spectral power persists in these regions in the supercontinuum at the fiber output because it is generated towards the end of the fiber. However, as noted above, the damage is observed to occur at a point where the pump light has not yet been spectrally broadened. We estimate the spectrum-weighted absorption coefficient of methane for just the pump spectrum to be around $\qty{0.04}{\per\m}$. From Eq.~\ref{eqn:dt}, this suggests that each $\qty{5}{\micro\joule}$ pump pulse contributes to a temperature rise of approximately $\Delta T \sim \qty{0.007}{\K}$ in the gas. Therefore, at a laser repetition rate of 10~kHz, it would take only about 3 seconds to exceed a temperature of \qty{200}{\degreeCelsius}, indicating that linear absorption is a plausible cause of the heating observed in the gas within the fiber.

However, as our estimate is oversimplified, the question remains: is the heating only caused by linear absorption of the gas or also by nonlinear absorption during the Raman interaction, as commonly observed in other experiments~\cite{beetar_thermal_2021}? To empirically answer this question we performed several additional experiments.

First, we tested using lower pulse energy ($\SI{350}{\nano\joule}$) but at higher repetition rates. This reduces the pulse peak power (and hence nonlinear effects), but maintains the average power. In this case, we could readily exceed \qty{50}{\kHz}, and sustain higher average powers through the fiber without observing damage. We reached up to \qty{600}{\kHz} for 20 minutes without damage---four times the average power of the $\SI{5}{\micro\joule}$ pulses at \qty{50}{\kHz}. Second, we tested the use of ethylene instead of methane. Ethylene does not exhibit absorption around the pump wavelength, unlike methane, but exhibits a somewhat similar Raman response. We still observed significant nonlinear absorption effects in ethylene. Initially, these observations led us to conclude that nonlinear absorption was the dominant factor. However, this conclusion was moderated by a third set of experiments, performed when pumping a methane-filled fiber with the second harmonic of our 1030~nm pump laser, at \qty{515}{\nm}. When pumping at \qty{515}{\nm} there is no linear absorption around the pump, even for methane, and hence the nonlinear contribution can be isolated. At $\SI{5}{\micro\joule}$ pump energy we scaled the repetition rate up to \qty{200}{\kHz} without observing damage. At lower energies (around $\SI{350}{\nano\joule}$), we reached up to \qty{2}{\MHz}, almost double the average power. The mixed results from these experiments lead us to conclude that both linear and nonlinear absorption play a role. It seems reasonable, however, to suggest that selecting a molecular gas that does not exhibit linear absorption in the spectral region of the supercontinuum formation is preferable, as is selecting a gas that can tolerate high temperatures before decomposing, while also selecting a gas that avoids the universal gain suppression effects previously observed in nitrogen~\cite{Sabbah:23}.

In summary, we have demonstrated ultra-flat multi-octave supercontinuum spanning from 350~nm up to 1700~nm using methane-filled antiresonant fiber, and observed that both modulation instability and Raman comb formation contribute to the supercontinuum evolution. Unlike the previously investigated case of nitrogen, gain suppression is avoided in methane. We investigated power scaling at higher pump repetition rates and reached up to 50~kHz, but higher powers and longer exposures at several tens of kHz rates caused fiber degradation. Our results suggest that both linear and nonlinear absorption play a role in methane, and repetition rate scaling, and hence power-scaling, will be limited, due to the tendency for methane to decompose at relatively low temperatures.\linebreak

\noindent\textbf{Acknowledgments.} We thank Martin Gebhardt for useful discussions.\linebreak

\noindent\textbf{Funding.} Royal Academy of Engineering (RF/202021/20/310, RF/202122/21/133). JCT is supported by a Chair in Emerging Technology from the Royal Academy of Engineering and by the Institution of Engineering and Technology (IET) through the IET A F Harvey Engineering Research Prize.\linebreak

\noindent\textbf{Disclosures.} The authors declare no conflicts of interest.\linebreak

\noindent\textbf{Data availability.} The data that support the findings of this study are available from the corresponding author upon reasonable request.\linebreak

% \the\columnwidth
% \the\textwidth
% Bibliography
\bibliography{MyLibrary}
% \bibliographyfullrefs{MyLibrary}

\end{document}